\documentclass[usenatbib]{mn2e}
\usepackage{mathptmx,amssymb,amsmath,natbib,fixltx2e}
\usepackage{graphics,graphicx,color,epstopdf,epsfig}

\include{defn}
\def\gtrsim{\lower 2pt \hbox{$\, \buildrel {\scriptstyle >}\over
{\scriptstyle \sim}\,$}}
\def\lesssim{\lower 2pt \hbox{$\, \buildrel {\scriptstyle <}\over
{\scriptstyle \sim}\,$}}

\begin{document}

\title{Catching jetted tidal disruption events early in millimetre}

\author[]{Qiang Yuan$^1$\thanks{E-mail:yuanq@pmo.ac.cn},
Q. Daniel Wang$^1$\thanks{E-mail:wqd@astro.umass.edu},
Wei-Hua Lei$^2$, He Gao$^3$ and Bing Zhang$^4$\\
$^1$Department of Astronomy, University of Massachusetts, 710 North
Pleasant St., Amherst, MA, 01003, USA\\
$^2$School of Physics, Huazhong University of Science and Technology,
Wuhan 430074, China\\
$^3$Department of Astronomy, Beijing Normal University, Beijing 100875, China\\
$^4$Department of Physics and Astronomy, University of Nevada Las Vegas,
NV 89154, USA}

\maketitle

\label{firstpage}

\begin{abstract}
Relativistic jets can form from at least some tidal disruption events
(TDEs) of (sub-)stellar objects around supermassive black holes.
%We show that millimeter (MM) observations can be a powerful tool to detect
%such jets and to probe their interplay with the circumnuclear media.
We detect the millimeter (MM) emission of IGR J12580+0134 --- the
nearest TDE known in the galaxy NGC 4845 at the distance of only 17
Mpc, based on Planck all-sky survey data. The data show significant
flux jumps after the event, followed by substantial declines, in all
six high frequency Planck bands from 100 GHz to 857 GHz. We further
show that the evolution of the MM flux densities are well consistent
with our model prediction from an off-axis jet, as was initially
suggested from radio and X-ray observations. This detection
represents the second TDE with MM detections; the other is Sw
J1644+57, an on-axis jetted TDE at redshift of 0.35. Using the on-
and off-axis jet models developed for these two TDEs as templates,
we estimate the detection potential of similar events with the Large
Millimeter Telescope (LMT) and the Atacama Large
Millimeter/submillimeter Array (ALMA). Assuming an exposure of one
hour, we find that the LMT (ALMA) can detect jetted TDEs up to
redshifts $z\sim1$ (2), for a typical disrupted star mass of $\sim1$
M$_\odot$. The detection rates of on- and off-axis TDEs can be as
high as $\sim0.6$ (13) and 10 (220) per year, respectively, for the
LMT (ALMA). We briefly discuss how such observations, together with
follow-up radio monitoring, may lead to major advances in
understanding the jetted TDEs themselves and the ambient environment
of the CNM.
%determining (1) the overall population of jetted TDEs, (2) the structure
%and energetics of the jets, (3) the density and magnetic field structures
%of the CNM, and (4) the origin of ultra high-energy cosmic rays.
\end{abstract}


\section{Introduction}

Supermassive black holes (SMBHs) are believed to be present at the
nuclei of all major galaxies. While only a small fraction of these
SMBHs are active, the properties of the silent majority could be
revealed by observations of tidal disruption events (TDEs) --- phenomena
in which stars or sub-stellar objects are tidally disrupted when they
pass close enough by SMBHs \citep[e.g.,][]{1988Natur.333..523R,
1989ApJ...346L..13E}. Part of the debris of a disrupted object would
be accreted onto the black hole, producing flaring X-ray and optical
emission with a typical $t^{-5/3}$ light curve which traces the fallback
rate of the stellar material \citep{1989IAUS..136..543P}. TDEs are expected
to occur every $10^3-10^5$ years for a typical galaxy \citep[e.g.,][]
{1999MNRAS.309..447M,2004ApJ...600..149W}. This rate could be
substantially enhanced (as high as once per a few years) in galaxy nuclei
with binary SMBHs \citep[e.g.,][]{2009ApJ...697L.149C,2011ApJ...729...13C}.

Energetic jets can be launched by a TDE. When they interact with the
circum-nuclear medium (CNM), high energy particle acceleration
occurs \citep[e.g.,][]{2006ApJ...645.1138C,2009ApJ...693..329F}.
Long-lasting, non-thermal radio emissions confirm that at least some
TDEs did launch relativistic jets. This catalogue includes Sw J1644+57
\citep{2011Sci...333..203B,2011Natur.476..421B,2011Natur.476..425Z,
2012ApJ...748...36B,2013ApJ...767..152Z}, Sw J2058+05
\citep{2012ApJ...753...77C}, Sw J1112-82 \citep{2015MNRAS.452.4297B},
IGR J12580+0134 \citep{2013A&A...552A..75N,2015ApJ...809..172I}, and
possibly ASASSN-14li\footnote{Note, however, \citet{2016ApJ...819L..25A}
interpreted this event as a non-relativistic outflow similar to supernova
ejecta.} \citep{2016Sci...351...62V}. The fraction of jetted TDEs is yet
unclear. \citet{2013ApJ...763...84B} searched for the late-time radio
emission from seven X-ray selected TDEs and found that two of them
might have radio counterparts, which implied that $\gtrsim10\%$ of the
X-ray-detected TDEs might have launched relativistic jets. However,
\citet{2013A&A...552A...5V} observed seven other TDE candidates, all of
which triggered within 10 years, but found that none of them had radio
emission up to $10\,\mu$Jy level, although most of these sources
are relatively distant ($z\approx0.14-0.37$). At present, the number
statistics is too small to allow for a reliable estimate of the jetted
fraction. Enlarging the jetted TDE sample is thus crucial to the
understanding of the jet formation (e.g., its dependence on SMBH spins;
\citealt{2011ApJ...740L..27L}).

TDE jets also provide unique diagnostics of the environment around SMBHs.
The jet emission reveals the CNM density and/or magnetic field (producing
Faraday rotation) profiles on sub-parsec scales, which can hardly be
probed in any other ways. Unlike $\gamma$-ray bursts (GRBs) whose time
scales are typically of the order of hours to days for afterglows, the
corresponding observable time scales of TDEs are of the order of months
to years. A long time scale allows for easy follow-up of such events,
especially for the early development of jets. Furthermore, the spatial
scales of TDEs are also large enough, allowing for spatially-resolved
studies for some nearby events (e.g., IGR J12580+0134 at $d\approx17$ Mpc).

Timely follow-up observations of TDE jets are very important for
probing their early evolution. However, early follow-ups are
strongly affected by the self-absorption of synchrotron emission in
radio \citep{2015ApJ...809..172I}. The infrared and optical emission
from jets is typically not expected to be bright and could suffer
serious confusion from galactic emission and/or extinction of the
interstellar medium, especially in the nuclear regions of the host
galaxies. Observing TDEs in (sub-)millimeter (MM) is thus optimal
for detecting the jets in the earliest stages. As the jets expand in
the CNM, the emission gradually becomes optically thin, first at
high frequencies and eventually down to the radio. A complete view
of the jet evolution from its earliest stages can potentially allow
us to understand the overall energetics of the jets, the CNM
environment, as well as the high energy particle acceleration. In
this work we assess this potential of observing TDE candidates
(which are supposed to be discovered typically in X-ray and optical
surveys\footnote{Radio surveys can also discover TDE-like transients
\citep{2015ApJ...803...36D,2015ApJ...806..224M}. However, those
observations are supposed to have relatively large cadence in
general. The self-absorption of the low frequency emission as
mentioned above is also a problem for the early monitoring.}) with
the MM facilities such as the Large Millimeter Telescope {\it
Alfonso Serrano} (LMT\footnote{http://www.lmtgtm.org/}) and the
Atacama Large Millimeter Array
(ALMA\footnote{http://www.almaobservatory.org/}), in addition to the
presentation of our data analysis results based on the existing
Planck all-sky survey.

This paper is organized as follows. In \S~2, after a brief
introduction on the current MM observations of TDEs, we present an
analysis of the Planck data on IGR J12580+0134. Taking Sw J1644+57
and IGR J12580+0134 as the templates of on- and off-axis jetted
TDEs\footnote{On-axis jets are those moving along the line-of-sight
(LOS) toward us. Specifically, they are defined to satisfy
$\theta_{\rm obs}\leq\theta_j$, where $\theta_{\rm obs}$ is the
angle between the jet axis and the LOS and $\theta_j$ is the jet
opening angle. Otherwise, they are off-axis jets.}, we describe the
modeling of the emission from TDE jets in \S~3. In \S~4 we study the
detectability and event rates of jetted TDEs with the LMT and ALMA.
We discuss possible physical insights of MM observations on TDEs and
related topics in \S~5, and finally summarize our work in \S~6.

\section{MM detections of TDEs with existing observations}

Up to now there are about 60 candidates of TDEs\footnote{See
http://astrocrash.net/resources/tde-catalogue/}, most of which were
discovered in X-ray and optical. A few of them were long-lasting radio
sources, suggesting the scenario of synchrotron emission from non-thermal
electrons accelerated at the jet-induced shocks in the CNM. In the MM
bands, only Sw J1644+57 ($z=0.35$) was detected with the Combined Array for
Research in Millimeter Astronomy (CARMA\footnote{https://www.mmarray.org/})
at 87 GHz and the Submillimeter Array
(SMA\footnote{http://www.cfa.harvard.edu/sma/}) at 200, 230 and 345 GHz
\citep{2012ApJ...748...36B}. The MM emission peaks at $t\lesssim10$ days
after the outburst of the TDE and has a flux $\sim20$ mJy at 87 GHz
\citep{2012ApJ...748...36B}. The combined MM and radio observations
extending to $\sim200$ days show clearly two components of the light curves
(see \S~3.1 and Fig. \ref{fig:Sw1644}), indicating that the jet is structured
\citep{2012ApJ...748...36B,2014ApJ...788...32W,2015ApJ...798...13L}.

%\begin{table*}
%\centering
%\caption{Planck counterparts of TDE IGR J12580+0134}
%\begin{tabular}{cccccc}
%\hline \hline
%$\nu$ & PCCS1 & PCCS2 \\
%(GHz) & 2009-08-12 $-$ 2010-11-27 & 2009-08-12 $-$ 2012-01-11 \\
%\hline
%353 & $<289$ mJy$^a$ & G306.77+64.37 ($504\pm61$ mJy; $1'.9^b$) \\
%545 & G306.73+64.40 ($1381\pm127$ mJy; $0'.3^b$) & G306.74+64.40 ($1637\pm105$ mJy; $0'.1^b$) \\
%857 & G306.72+64.40 ($5784\pm157$ mJy; $0'.6^b$) & G306.74+64.39 ($6178\pm186$ mJy; $0'.5^b$) \\
%\hline
%\hline
%\end{tabular}\\
%Note: $^a$Corresponding to the minimum flux density of the catalogue
%at $|b|>30^{\circ}$ after excluding the faintest $10\%$ of sources
%\citep{2014A&A...571A..28P}.
%$^b$Offset from the central position of the host galaxy NGC 4845,
%$(l,b)=(306^\circ.7413,64^\circ.3986).$
%\label{table:pccs}
%\end{table*}

\begin{figure*}
\centering
\includegraphics[width=1.15\columnwidth]{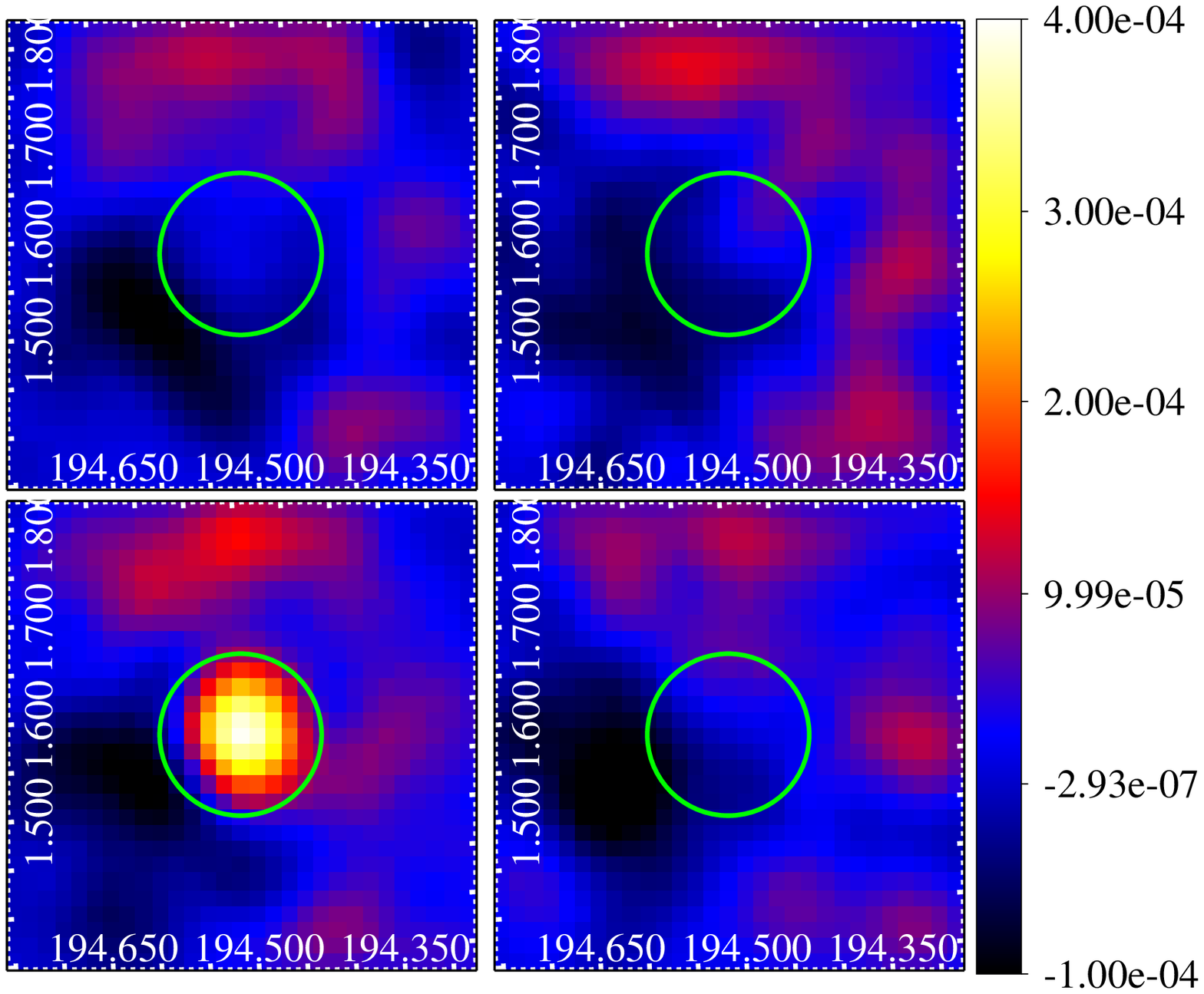}
\hspace*{-3cm}
\includegraphics[width=1.28\columnwidth]{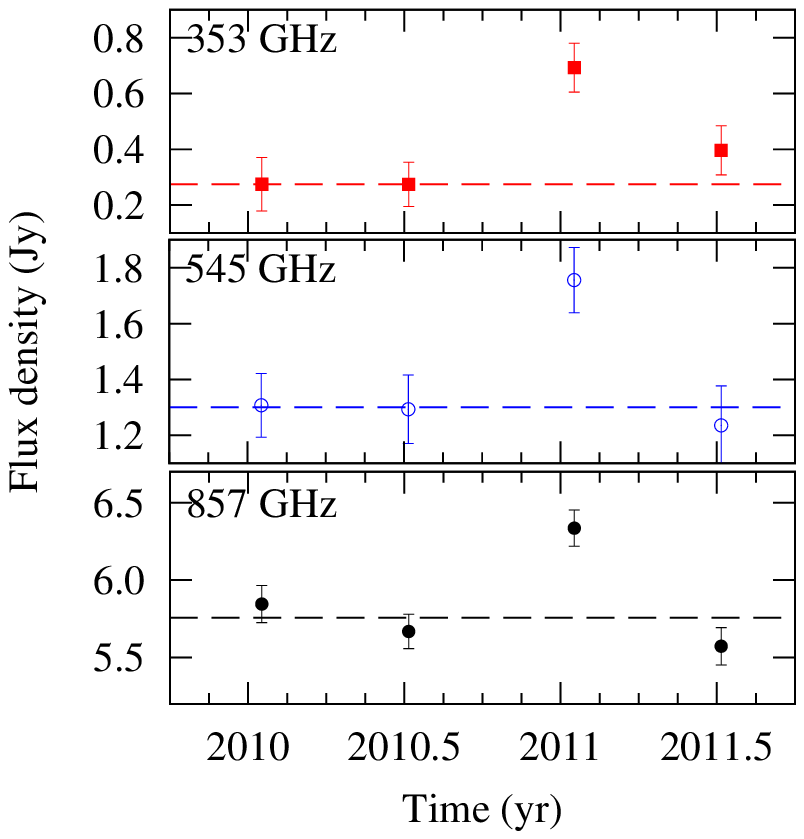}
\caption{Left panel: $30'\times30'$ minimaps (smoothed with a
Gaussian kernel with $\sigma=2'$) of Planck observations of NGC 4845
at 217 GHz, at $T_1$ (top-left), $T_2$ (top-right), $T_3$
(top-left), and $T_4$ (top-right). The green circle outlines a $10'$
diameter region centered at NGC 4845 in each sub-panel. The units of
the colorbar are ${\rm K_{CMB}}$. Right panel: the four epoch flux
densities of NGC 4845 at the 353 (top), 545 (middle), and 857
(bottom) GHz frequencies. The dashed line in each sub-panel shows
the average flux densities of the first two epochs, which represents
the emission of the galaxy. } \label{fig:planck}
\end{figure*}

\begin{table*}
\centering \caption{Flux densities of the Planck observations of NGC
4845.}
\begin{tabular}{ccccccc}
\hline \hline
$\nu$ & 2010-01-16 ($T_1$) & 2010-07-05 ($T_2$) & 2011-01-16 ($T_3$) & 2011-07-05 ($T_4$) & TDE at $T_3$\\
(GHz) &   (mJy)    &   (mJy)    &   (mJy)    &   (mJy)   &    (mJy)   \\
\hline
30 & --- & --- & --- & --- & $<427^a$\\
44 & --- & --- & --- & --- & $<692^a$\\
70 & --- & --- & --- & --- & $<501^a$\\
100 & --- & --- & $640\pm103$ & --- & $635\pm103^b$\\
143 & --- & --- & $741\pm37$ & --- & $726\pm37^b$\\
217 & --- & --- & $675\pm42$ & --- & $615\pm42^b$\\
353 & $275\pm96$ & $274\pm79$ & $693\pm87$ & $396\pm88$ & $419\pm107^c$\\
545 & $1307\pm114$ & $1293\pm122$ & $1756\pm117$ & $1236\pm142$ & $456\pm144^c$ \\
857 & $5846\pm120$ & $5668\pm111$ & $6336\pm117$ & $5573\pm120$ & $579\pm143^c$ \\
\hline
\hline
\end{tabular}\\
Notes: $^a$Minimum flux density at $90\%$ completeness
\citep{2015arXiv150702058P}.
$^b$Flux density of IGR J12580+0134 at $T_3$ after subtracting the emission
of the galaxy which is assumed to follow the power-law extrapolation of
those (mean of $T_1$ and $T_2$) at 353, 545, and 857 GHz.
$^c$Inferred flux density of IGR J12580+0134 at $T_3$ after subtracting
the mean emission of $T_1$ and $T_2$.
\label{table:planck}
\end{table*}

We note that Planck surveyed the MM sky for 8 times with the low
frequency instrument (LFI; at 30, 44, and 70 GHz) and 5 times with
the high frequency instrument (HFI; at 100, 143, 217, 353, 545, and
857 GHz) during its four years' operation. Two Planck catalogues of
compact sources (PCCS), with the first one (PCCS1) covering the
period from August 12, 2009 to November 27, 2010, and PCCS2 covering
from August 12, 2009 to January 11, 2012 (for the HFI), have been
released by the Planck team
\citep{2014A&A...571A..28P,2015arXiv150702058P}. We thus conduct a
search for potential counterparts for the TDE candidates which were
detected from the beginning of 2009 to the end of 2011 in the PCCS1
and PCCS2 catalogues. About 20 TDE candidates are searched, and we
find a counterpart of the TDE IGR J12580+0134, discovered by
INTEGRAL \citep{2013A&A...552A..75N} in December, 2010, in the
nearby galaxy NGC 4845. This counterpart is represented by a
variable source positionally coincident (within the beam full widths
at half maximum (FWHM), which are $\sim5-10'$ for the HFI;
\citealt{2014A&A...571A..28P}) with IGR J12580+0134. The fluxes seen
in PCCS2 are systematically higher than those in PCCS1, suggesting
the emission from the TDE.

To quantify the MM emission of the TDE, we have systematically
analyzed the Planck survey data. Table \ref{table:planck} summarizes
the detected emission in the four epochs covered by the HFI. The
left-hand panel of Fig. \ref{fig:planck} shows our extracted 217 GHz
minimaps\footnote{http://irsa.ipac.caltech.edu/applications/planck/}
centered at NGC 4845 for these four epochs. These maps have been
smoothed with a Gaussian kernel with $\sigma$ of 2'. A substantial
brightening at the source location can be seen at $T_3$ (bottom-left
sub-panel), just after the outburst of the TDE. We estimate the flux
density of the source in the raw imaging data of each map using a
Gaussian fit photometry method
\citep{2014A&A...571A..28P,2015arXiv150702058P}. Since the galaxy is
unresolved by Planck, we employ a symmetric Gaussian function with
width $\sigma$ fixed at FWHM/2.355, plus a constant background, to
fit the image. The FWHM values of each image is adopted from
\citet{2014A&A...571A..28P}. The normalization of the noise is
obtained by setting the best-fit reduced $\chi^2=1$, and the
uncertainty of the flux density is computed from the curvature of
the $\chi^2$ \citep{2014A&A...571A..28P,2015arXiv150702058P}. The
results are given in Table \ref{table:planck}. No significant source
is found in the LFI observations, and the minimum flux densities
corresponding to $90\%$ completeness of the PCCS2 catalogue are
given. For the HFI bands, the galaxy is detected at all four epochs
at 353, 545, and 857 GHz. The flux densities in these three bands
are shown in the right-hand panel of Fig. \ref{fig:planck}, as well
as in Table \ref{table:planck}. We see that the flux densities of
the galaxy increase significantly at $T_3$ for all three bands,
compared to those at $T_1$ and $T_2$. The emission droped to the
average level at $T_4$, except for the 353 GHz band which was still
brighter than the average. Taking the mean flux densities of $T_1$
and $T_2$ as the emission of the galaxy, we can derive the flux
densities of TDE IGR J12580+0134 at $T_3$, which are given by the
last column of Table \ref{table:planck}. For the 100, 143, and 217
GHz bands, the emission is detected only at $T_3$. The flux
densities of the TDE are then derived through subtracting the
extrapolated galaxy emission assuming a power-law spectrum according
to the higher frequency results (mean of that of $T_1$ and $T_2$).

\section{Models for jetted TDEs}

Here we provide an overview of the afterglow emission modeling of jetted
TDEs. Taking Sw J1644+57 and IGR J12580+0134 as examples of on- and off-axis
jetted TDEs, we demonstrate the modeling of the multi-band and multi-epoch
data. The model is somehow simple, with one-dimensional (radial) evolution
of the jets. The two-dimensional model with the lateral expansion
\citep{2015MNRAS.450.2824M,2016arXiv160508437G} would describe the jet
dynamics more precisely. However, for the purpose of this work and/or the
data quality for these objects, the one-dimensional model is expected to
be adequate.

\subsection{On-axis TDEs}

\begin{figure*}
\centering
\includegraphics[width=\columnwidth]{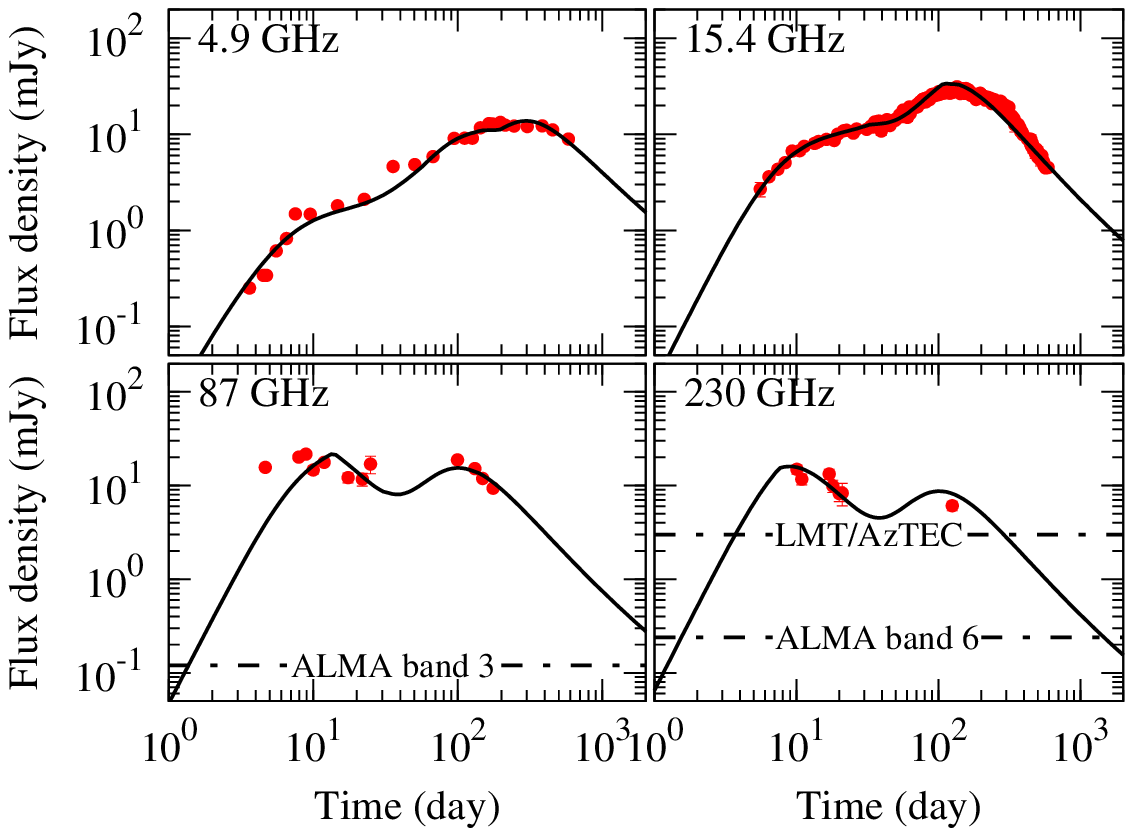}
\includegraphics[width=\columnwidth]{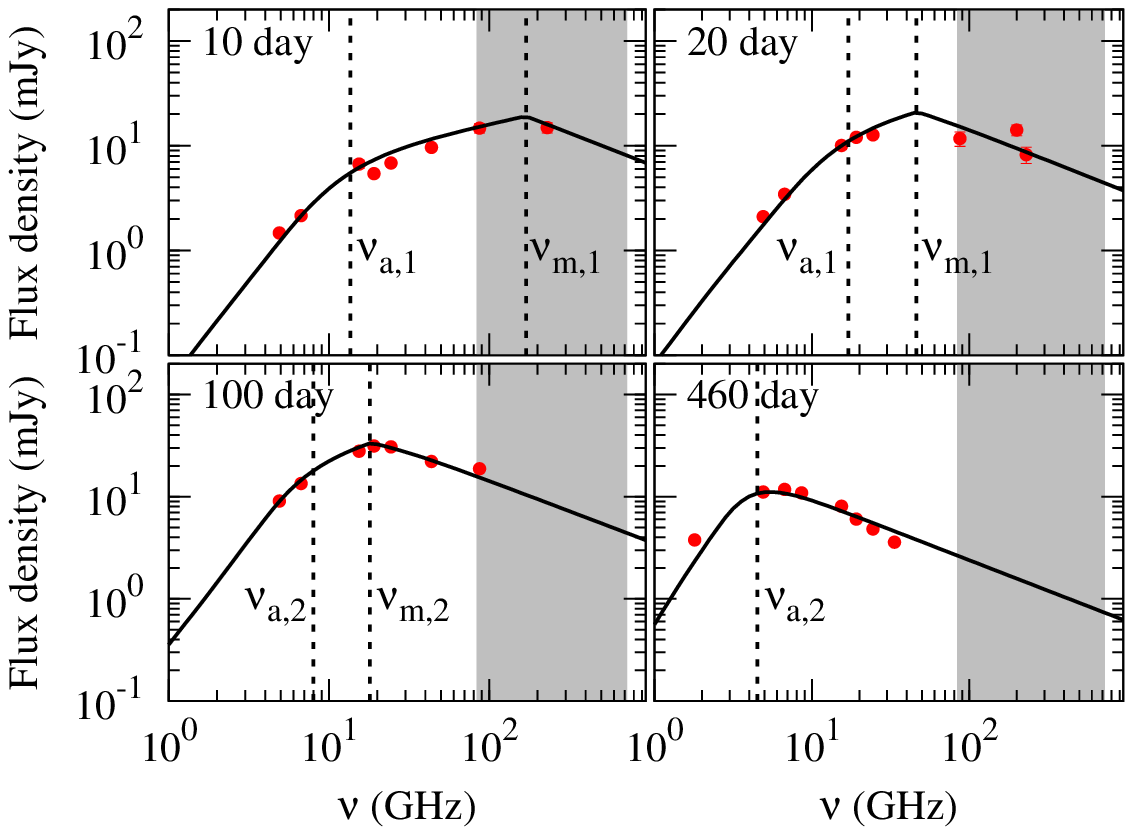}
\caption{Left: light curves of the on-axis TDE Sw J1644+57 at 4.9, 15.4,
87, and 230 GHz, respectively. The solid lines show the predictions of
the two-component jet model \citep{2014ApJ...788...32W}, compared with
the observational data \citep{2012ApJ...748...36B,2013ApJ...767..152Z}.
Horizontal lines show the sensitivities of the LMT and ALMA at
different bands (Table \ref{table:sen}). Right: SEDs of Sw J1644+57 at
10, 20, 100, and 460 day, respectively. Vertical dotted lines represent
the typical frequencies $\nu_a$ (the self-absorption frequency) and
$\nu_m$ (the frequency corresponding to the minimum injection energy
of electrons). The subscription ``1'' (``2'') refers to the inner-fast
(outer-slow) jet which dominates the emission at early (late) stages.
}
\label{fig:Sw1644}
\end{figure*}

\begin{table*}
\begin{center}
\caption{Model parameters from the radio data fitting \label{table:para}}
\begin{tabular}{ccccccccccc}
\hline\noalign{\smallskip}
  & $\Delta t$ (day) &  $n_{18}$ (${\rm cm}^{-3}$) & $k$ & $\theta_{\rm obs}$ (deg) & $E_{50}^a$ & $\Gamma_j$ &
  $\theta_j$ (deg) & $p$ & $\epsilon_e^b$ & $\epsilon_B^b$ \\
\hline\noalign{\smallskip}
Sw J1644+57 (inner) & 0.25$^c$ & 0.05 & 0.0 & 1.4 & 9000  & 8.7 & 1.7 & 2.20 & 0.13 & 0.13 \\
Sw J1644+57 (outer) & 0.25$^c$ & 0.05 & 0.0 & 1.4 & 9000  & 3.6 & 4.5 & 2.20 & 0.25 & 0.31 \\
\hline
IGR J12580+0134 & 30$^d$ & 8.8 & --- & 35 & 72   & 4.6 & 9.5 & 1.80 & 0.25 & 0.25 \\
\noalign{\smallskip}\hline
\end{tabular}\\
Notes: Columns from left to right are: source name, jet launching time,
CNM density at $10^{18}$ cm, power-law index of the density profile, viewing
angle, kinetic energy of the jet, initial Lorentz factor of the jet,
opening angle of the jet, spectral index of accelerated electrons, electron
energy density fraction, and magnetic field energy density fraction. \\
$^a$In $10^{50}$ erg s$^{-1}$.
$^b$Fraction of the ejecta kinetic energy assigning to accelerated
electrons or the magnetic field.
$^c$Relative to March 25.5, 2011.
$^d$Relative to December 12, 2010.
\end{center}
\end{table*}

The physical picture of on-axis jetted TDEs is similar to that of GRBs.
The central engine, the tidal disruption of stellar objects by SMBHs,
powers relativistic jets, which then propagate in the CNM. The internal
dissipation within the jets, probably together with emission from the
accretion disk and/or its corona, could be responsible for the early
X-ray/UV/optical emission, which corresponds to the prompt emission of
GRBs. The jet-CNM interaction generates shocks to accelerate electrons
(or even cosmic rays) which radiate via synchrotron and/or inverse
Compton scattering, giving rise to the so-called afterglow emission.
The evolution of a jet during the propagation in the CNM includes
roughly three stages. The jet first undergoes a coasting phase with a
nearly constant speed. Then it starts to decelerate when the mass of
the CNM swept by the forward shock is comparable to $M_{\rm ej}/\Gamma_j$,
where $\Gamma_j$ and $M_{\rm ej}$ are the initial Lorentz factor and mass
of the ejecta. Finally the blastwave enters the Newtonian phase when
the rest mass equivalent energy of the swept CNM is comparable to the
initial kinetic energy of the jets. The dynamics of the jet may then be
described by a set of hydrodynamical equations \citep{2000ApJ...543...90H}.
We here adopt a numerical approach to describe the jet dynamics, and
follows \citet{1998ApJ...497L..17S} (see \citealt{2013NewAR..57..141G}
for a more comprehensive version) to calculate the synchrotron emission
from the accelerated electrons. We have a total of 10 parameters to
describe the jet dynamics, radiation, and the CNM density structure,
as shown in Table \ref{table:para}.

A two-component jet model has been developed by
\citet{2014ApJ...788...32W}, to explain the complex multi-wavelength
light curves of Sw J1644+57. We use a similar but improved version of
the model to fit the multi-wavelength and multi-epoch data
\citep{2012ApJ...748...36B,2013ApJ...767..152Z}. Instead of a constant
density of the CNM, we assume a power-law distribution $n(r)=n_{\rm 18}
(r/{10^{18}\,{\rm cm}})^{-k}$. Furthermore, we introduce an additional
parameter $\Delta t$, after the outburst of the event, to characterize
the zero point of the jet evolution. Fig. \ref{fig:Sw1644} shows the
model light curves (left) and the spectral energy distributions (SED;
right) that fit the observational data. The two bumps in the light curves
(left-hand panels of Fig. \ref{fig:Sw1644}) represent the contributions
from the inner-fast and the outer-slow jets, which dominate the emission
at early and late stages, respectively. We find a small difference (with
$\Delta t=0.25$ days) of the jet launching time relative to the $\gamma$-ray
outburst time of Sw J1644+57, March 25.5, 2011 \citep{2011Natur.476..425Z}.
The fitting suggests $k\sim0$, indicating that the jets propagate in a
uniform, low density environment. However, we note from Fig.
\ref{fig:Sw1644} that this model does not reproduce the light curves and
SEDs well enough. Further refinement of the model is necessary, possibly
including the spectral evolution of accelerated electrons.

The model parameters are given in Table \ref{table:para}. These parameters
are somewhat different from those in \citet{2014ApJ...788...32W}, because
of the likelihood fitting technique adopted in this work. Note, however,
these parameters are only a set of representative ones which can
describe the data, because of strong degeneracies of the model
parameters. A quantitative assessment of the confidence regions of
the parameters is non-trivial due to the systematic uncertainties of
the radio observations (from e.g., calibration and scintillation),
and beyond the purpose of this study. Furthermore, the model parameters
from the two-dimensional model are expected to be different from those
from the one-dimensional one adopted here \citep{2015MNRAS.450.2824M}.

\begin{figure*}
\centering
\includegraphics[width=\columnwidth]{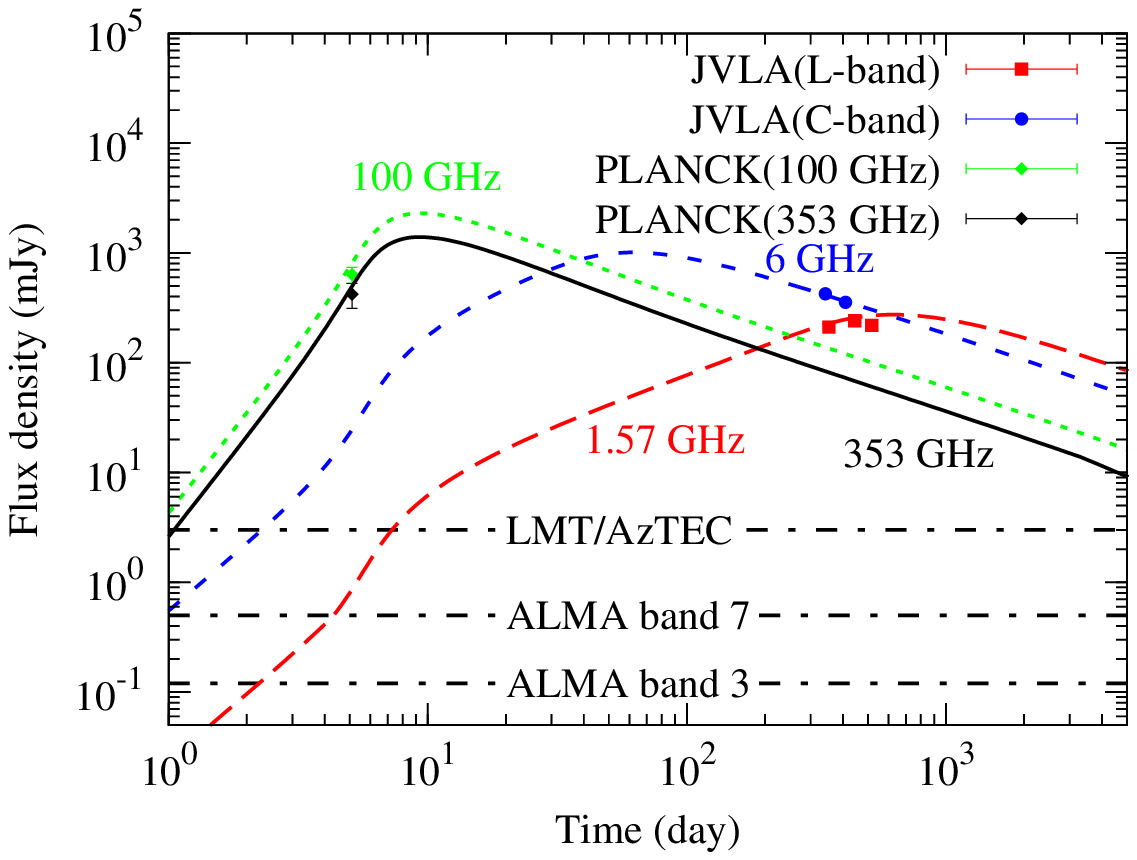}
\includegraphics[width=\columnwidth]{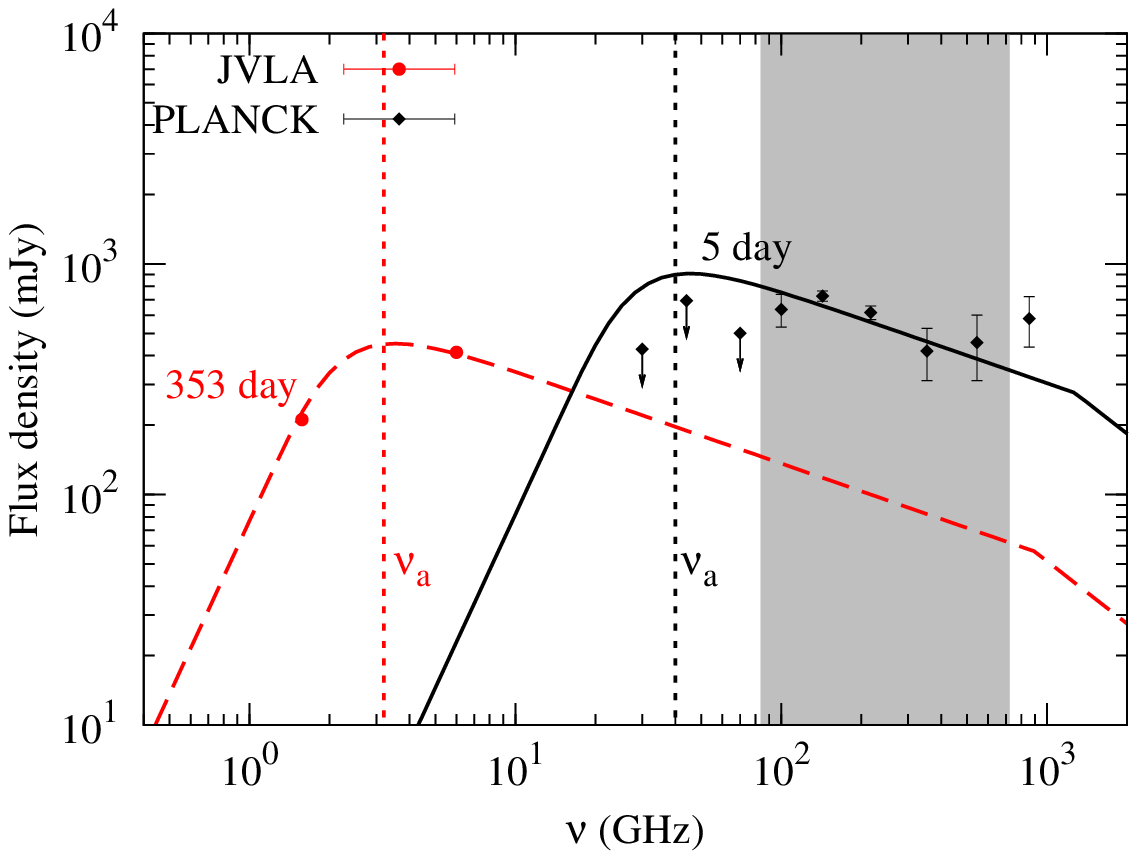}
\caption{Left: model light curves of the off-axis TDE IGR
J12580+0134 at 1.57, 6, 100, and 353 GHz. Right: SED of IGR
J12580+0134 at 5 and 353 days. The radio data are from
\citet{2015ApJ...809..172I}, and the MM data are from Planck
analyzed in this work. Lines show the fitting results of the jet-CNM
interaction model. } \label{fig:NGC4845}
\end{figure*}

\subsection{Off-axis TDEs}

A TDE seen off-axis should be substantially less luminous than if observed
on-axis, due to the lack of significant Doppler boosting\footnote{
\citet{2015MNRAS.453..157P} showed that for a wide range of parameters,
the jet power is usually higher by orders of magnitudes than the disk
power. Therefore, the Doppler (de-)boosting is relevant for TDEs with
relativistic jets.}. The ratio of the off-axis to on-axis Doppler factors
is \citep{2002ApJ...570L..61G}
\begin{equation}
a_{\rm off}=\frac{{\mathcal D}_{\rm off}}{{\mathcal D}_{\rm on}}=
\frac{1-\beta}{1-\beta \cos \psi},
\label{eq:aoff}
\end{equation}
where $\beta$ is the velocity of a jet in units of the light speed $c$,
and $\psi=\max(\theta_{\rm obs}-\theta_j,\,0)$ is the angle between
the jet moving direction and the LOS. The bolometric luminosity scales
as\footnote{Note that the scaling relation here (and below for flux)
is valid for ``point source approximation'', i.e., $\theta_{\rm obs}
\gg\theta_j$. This assumption will underestimate the off-axis
contribution, making the event rate estimate in \S~4 more conservative.}
$L^{\rm off}\sim a_{\rm off}^4L^{\rm on}$. Typically for $\Gamma\sim$
a few (e.g., 4) and $\psi\sim\pi/4$, we have $a_{\rm off}\sim0.1$.
The factor $a_{\rm off}$ mainly takes effects at early time when
the jet is relativistic. With the deceleration of the jet, $a_{\rm off}$
approaches 1 asymptotically, and the difference between on- and
off-axis viewing angles becomes smaller. This can explain the $\sim7000$
times difference of the X-ray luminosities (after correcting the mass
difference of the disrupted objects) between Sw J1644+57 and IGR
J12580+0134 \citep{2016ApJ...816...20L}. The late time (e.g., $t\sim100$
day) radio luminosities of these two events, after correcting the mass
difference of the disrupted objects, are comparable with each other
when no significant effect from $a_{\rm off}$ applies.

The same model for on-axis TDEs can be applied to off-axis ones,
with proper modifications of the viewing angle effect, e.g.,
$\nu^{\rm off} =a_{\rm off}\nu^{\rm on}$, $t^{\rm off}=t^{\rm
on}/a_{\rm off}$, $F_{\nu}^{\rm off}(t)=a_{\rm off}^3F_{\nu/a_{\rm
off}}^{\rm on} (a_{\rm off}\,t)$. Since the data of IGR J12580+0134
are not good enough to constrain the CNM density profile, we assume
a constant density here. Fig. \ref{fig:NGC4845} shows the light
curves (left) and SEDs (right) for the model fitting results,
compared with the radio data by the Jansky Very Large Array
\citep[JVLA;][]{2015ApJ...809..172I} and the Planck data obtained in
\S~2. It is shown that the Planck HFI data are reasonably consistent
with the model at early time of the outburst. However, the Planck
LFI data may suggest a harder spectrum than that inferred from the
JVLA data, which implies a possible spectral evolution of
accelerated electrons in the TDE jets. The non-detection of the
source in the Planck LFI data may also partly be due to the optical
thickness of the emission at the early time. We further find that
the zero point of the jet launching is about 30 days after the first
X-ray detection on December 12, 2010 and about 12 days before the
X-ray peak time \citep{2013A&A...552A..75N}, which suggests a
non-jet origin of the X-ray emission \citep[see
also][]{2016ApJ...816...20L}.

Still the model parameters, shown in Table \ref{table:para}, just represent
one potential combination which can describe the data. The one-dimensional
model may also lead to differences compared with the two-dimensional case
\citep{2015MNRAS.450.2824M}. This is more severe for IGR J12580+0134
than that of Sw J1644+57, because the model is less constrained by the
limited data points. Since larger uncertainties of the emission of
off-axis jetted TDEs may come from the scaling of $a_{\rm off}$ factor
and the mass of the disrupted object \citep[about $10^2$ times smaller
for IGR J12580+0134 than a typical star;][]{2013A&A...552A..75N}, we do
not explore more detailed and complicated modeling in this work, but
take a few different scaling relations between $L^{\rm off}$ and
$L^{\rm on}$ to account for potential uncertainties instead.

\section{Potential capabilities of existing MM observing facilities}

\subsection{Facilities}

We take the LMT and ALMA as representives of single-dish telescopes
and telescope arrays for the discussion. Other facilities such as the James
Clerk Maxwell telescope (JCMT\footnote{http://www.eaobservatory.org/jcmt/}),
SMA, and CARMA, are expected to be within the capability ranges of the LMT
and ALMA. The LMT and ALMA, with latitudes $19^{\circ}$N and $23^{\circ}$S,
respectively, will cover the northern and southern sky complimentarily.
Furthermore, a single-dish telescope may be sufficient to observe a
relatively bright source and more flexible for a quick check of an event's
brightness, as well as for potential subsequent monitoring, while the
telescope array may be well suitable for detecting faint sources and
for high-resolution imaging of some nearby sources to resolve the jet
structures.

\subsubsection{LMT}

The LMT is a single-dish MM telescope, located on the summit of Volcan
Sierra Negra, Mexico at an altitude of 4600 m above the sea level
\citep{2010SPIE.7733E..12H}. It is designed for astronomical observations
in the wavelength range of $0.85-4$ mm. The LMT is currently operating
with the finished 32-meter diameter aperture and will soon have the
full 50-meter diameter. The corresponding angular resolution is $\sim8''$
currently (will finally be $\sim 5''$), and the pointing accuracy is
$\sim1''$. The two commissioned instruments are the Redshift Search
Receiver (RSR), a wide-band 3 mm spectrometer, and the Astronomical
Thermal Emission Camera (AzTEC), a 1.1 mm continuum camera. The sensitivity
of the AzTEC instrument is about $0.6$ mJy root-mean-square (RMS) for an
hour exposure under the small map mode, which is designed to map a 1.5
arcmin diameter region with uniform sensitivity (see Table \ref{table:sen}).

\subsubsection{ALMA}

The ALMA is located at the Atacama desert in Chile with an altitude of
about 5000 meters above the sea level. The dryness and high altitude
makes Atacama well suited for MM observations. The ALMA is a telescope
array consisting of 66 antennas when completed, 54 of them with 12-meter
diameter dishes and 12 smaller ones with 7-meter diameter. After the
full installation, the ALMA will cover 10 frequency bands of the (sub-)MM
windows from $\sim30$ GHz (10 mm) to THz (0.3 mm). Currently the ALMA
operates in part as the Early Science phase\footnote{See
https://science.nrao.edu/facilities/alma/earlyscience}, with more than
sixteen 12-meter antennas, yielding sensitivites of $\sim10\%$ of the
full array. The bands 3, 6, 7 and 9 are available, covering frequencies
from 84 GHz to 720 GHz (see Table \ref{table:sen}). The maximum
baseline is about 250 m, resulting in a maximum angular resolution
of $\sim2.5''$ in band 3 and $0.4''$ in band 9. The continuum sensitivity
is about $0.1-3.6$ mJy, for an hour exposure and a $5\sigma$ significance,
as shown in Table \ref{table:sen} \citep{2012A&A...538A..44D}.

\begin{table}
\centering
\caption{Typical (sub-)MM observational bands and sensitivities}
\begin{tabular}{cccccc}
\hline \hline
    & $\nu$ & $\lambda$ & sensitivity$^a$ \\
    & (GHz) & (mm) & (mJy) \\
\hline
LMT/AzTEC & 273 & 1.1 & 3.0$^b$ \\
ALMA band 3 & $84-116$ & $2.58-3.56$ & 0.12$^c$ \\
ALMA band 6 & $211-275$ & $1.09-1.42$ & 0.24$^c$ \\
ALMA band 7 & $275-373$ & $0.80-1.09$ & 0.50$^c$ \\
ALMA band 9 & $602-720$ & $0.42-0.50$ & 3.60$^c$ \\
\hline
\hline
\end{tabular}
\label{table:sen}\\
Notes: $^a$The sensitivity is defined as 5 times of the achievable
RMS for 1-hour observations. $^b$http://www.lmtgtm.org/?page\_id=713 .
$^c$For the early science operation of the ALMA \citep{2012A&A...538A..44D}.
\end{table}

\subsection{Detectability}

Figs. \ref{fig:Sw1644} and \ref{fig:NGC4845} compare the sensitivities
of the LMT and ALMA to the detected/expected fluxes of Sw J1644+57
and IGR J12580+0134. Both sources were bright enough to be detectable
at their peak time by the LMT and ALMA, and could still be detectable
years (for the LMT) or even ten years (for the ALMA) after their
outbursts. Note, however, the contamination from the galactic nuclei
may be important for an observation at a late stage when the flux has
decreased significantly, especially for a single-dish telescope.
Considering the peak fluxes, we find that the LMT (ALMA) can detect
Sw J1644+57-like events up to redshifts of $z\sim0.8$ (2.4), and IGR
J12580+0134-like events up to $z\sim0.1$ (0.35). If the disrupted
object mass of an IGR J12580+0134-like event is as massive as an ordinary
star ($\sim$M$_{\odot}$), such off-axis TDEs could be detectable up
to $z\sim0.7$ (2.2).

The wavelength coverages of the LMT and ALMA are shown with shaded regions
in the right-hand panels of Figs. \ref{fig:Sw1644} and \ref{fig:NGC4845}.
While the low frequency spectra are modifided by the self-absorption
and the minimum energy of accelerated electrons, the (sub-)MM observations
are weakly affected by these effects, and hence can directly measure the
spectral index of electrons. The MM emission in the early time may also
trace the evolution of $\nu_m$, which is helpful to understanding the
acceleration of electrons in the shocks.

\subsection{Detection rate}

The event rate density of Sw J1644+57-like (on-axis) jetted TDEs can be
estimated as \citep{2015ApJ...812...33S}
\begin{equation}
\rho^{\rm on}(L^{\rm on}_{p,{\rm bol}},z)=\frac{\rho^{\rm on}_0}{L_0^{\rm
on}}\left(\frac{L^{\rm on}_{p,{\rm bol}}}{L_0^{\rm on}}\right)^{-2.0}
\times f(z),\label{eq:on}
\end{equation}
where $L^{\rm on}_{p,{\rm bol}}$ is the peak bolometric luminosity in the
rest-frame $1-10^4$ keV range, $\rho_0\approx0.03$ Gpc$^{-3}$~yr$^{-1}$
is the local event rate density for source luminosities
above $L_0^{\rm on}=10^{48}$~erg~s$^{-1}$, and $f(z)$ gives the redshift
dependence of TDEs based on the semi-empirical models of the SMBH mass
density distribution
\citep{2013MNRAS.428..421S,2015ApJ...812...33S}
\begin{equation}
f(z)=\left[(1+z)^{0.2\eta}+\left(\frac{1+z}{1.43}\right)^{-3.2\eta}
+\left(\frac{1+z}{2.66}\right)^{-7.0\eta}\right]^{1/\eta},
\end{equation}
with $\eta=-2$.

We generalize Eq. (\ref{eq:on}) to include off-axis jetted TDEs as
\begin{equation}
\rho(L^{\rm off},z)=\epsilon^{-1}\times\rho^{\rm on}(L^{\rm on},z)
\cdot\frac{dL^{\rm on}}{dL^{\rm off}},
\label{eq:off}
\end{equation}
where we have ignored the subscript ``$p,{\rm bol}$'' of the luminosity
and $\epsilon=(1-\cos\theta_j)/(1-\cos\theta_{\rm obs}^{\rm max})$ is
the sky coverage fraction of the jets. In this work we assume
$\theta_j\sim15^{\circ}$. By definition we have $\theta_{\rm obs}^{\rm
max}=\theta_j$ (hence $\epsilon^{-1}=1$) for on-axis sources, and
$\theta_{\rm obs}^{\rm max}=\pi/2$ (hence $\epsilon^{-1}\sim30$) for off-axis
sources. For on-axis TDEs, $a_{\rm off}=1$, Eq. (\ref{eq:off}) recovers Eq.
(\ref{eq:on}). More generally Eq. (\ref{eq:off}) should be averaged over
$a_{\rm off}$ which is a function of the viewing angle $\theta_{\rm obs}$
and the Lorentz factor $\Gamma$. Because there are no good constraints on
the distribution of $\Gamma$, and hence $a_{\rm off}$, we assume a typical
value of $a_{\rm off}=0.1$ for the discussion of off-axis sources except
in Table \ref{table:num} where other values of $a_{\rm off}$ are considered
additionally. Fixing the value of $a_{\rm off}$ also makes it possible to
estimate the detectability of off-axis TDEs by MM facilities by re-scaling
the template source, IGR J12580+0134. However, one should keep in mind that
this assumption may be an over-simplification.

\begin{figure*}
\centering
\includegraphics[width=\columnwidth]{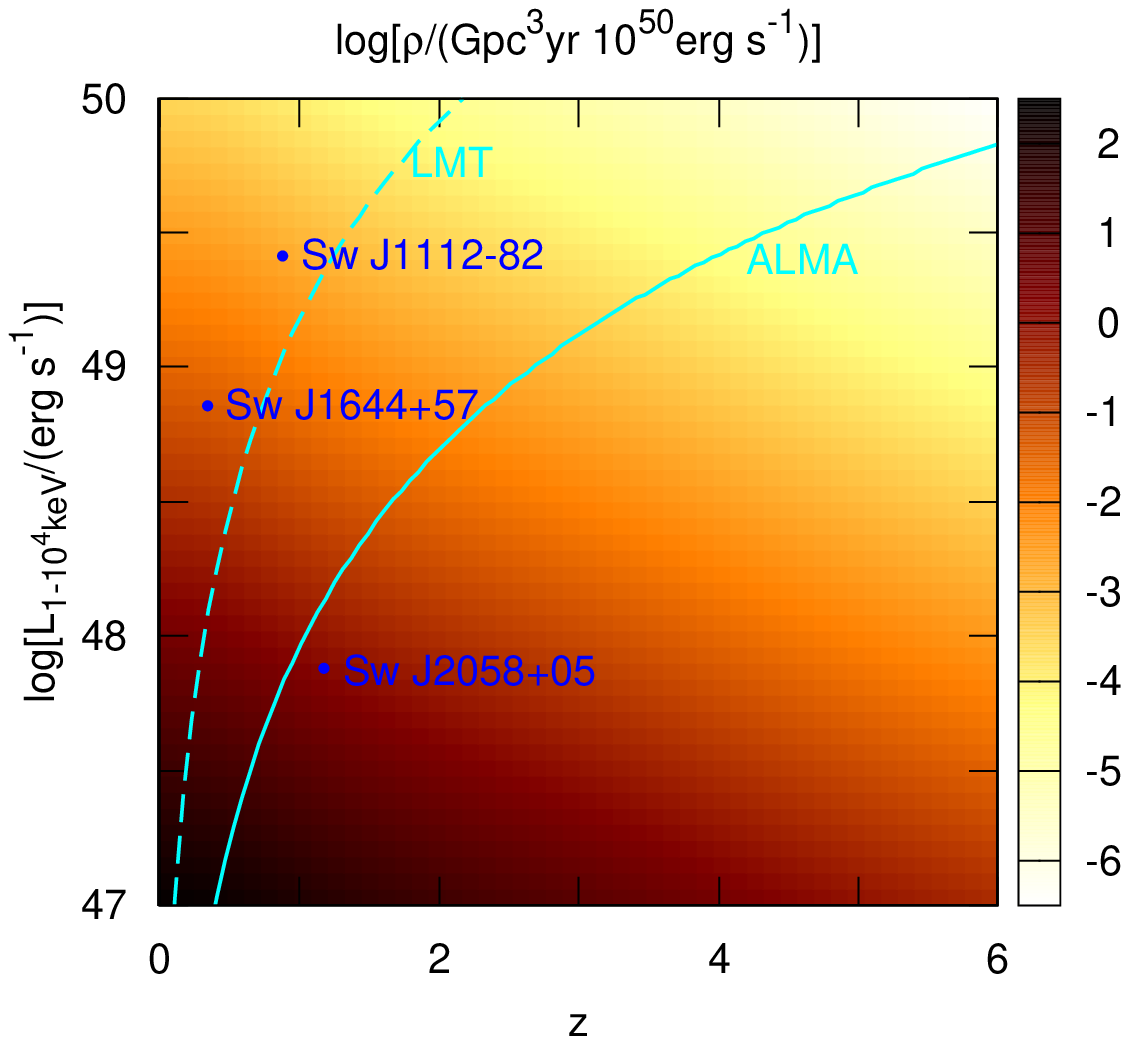}
\includegraphics[width=\columnwidth]{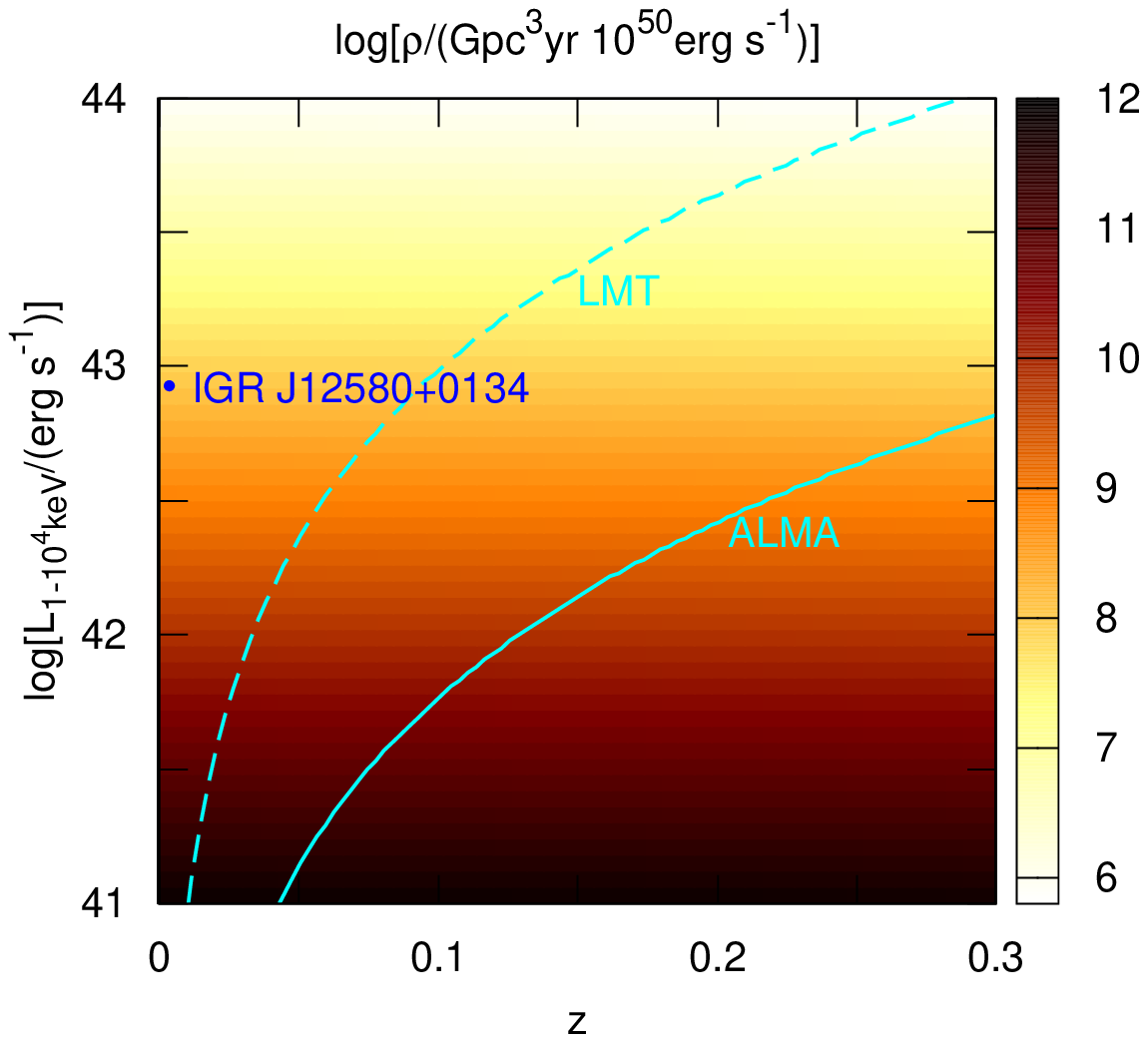}
\caption{Event rate density $\rho(L,z)$ of on-axis (left) and off-axis
(right; $a_{\rm off}^4=10^{-4}$) jetted TDEs as a function of redshift and
$1-10^4$ keV bolometric luminosity. Curves show the sensitivities of the
LMT and ALMA, above which TDEs are detectable, assuming that they can be
characterized by the models of energetical scaling of Sw J1644+57 (for
the on-axis type) or IGR J12580+0134 (for the off-axis type).
}
\label{fig:evrate}
\end{figure*}

Given a specific detection threshold, the detectable event rate is obtained
through integrating the event rate density over the luminosity and the
cosmological volume \citep{2015ApJ...812...33S}
\begin{equation}
\frac{dN}{dt}=\frac{\Omega}{4\pi}\int dL\int_0^{z_{\rm max}(L)}dz\,
\frac{\rho(L,z)}{1+z}\frac{dV}{dz},
\label{eq:num}
\end{equation}
where
\begin{equation}
\frac{dV}{dz}=\frac{c}{H_0}\frac{4\pi d_L^2}{(1+z)^2[\Omega_M(1+z)^3
+\Omega_{\Lambda}]^{1/2}}
\end{equation}
is the cosmological specific comoving volume (assuming the standard
$\Lambda$CDM cosmology) with $d_L$ the luminosity distance, $\Omega/4\pi$
is the fraction of the sky covered by the instrument, and $z_{\rm max}(L)$
is the maximum redshift that a source with luminosity $L$ can be detected.
In this work we assume $\Omega/4\pi\approx1/2$ for both the LMT and the
ALMA. The maximum redshift can be obtained through a scaling of the
template source as
\begin{equation}
d_L^2(z_{\rm max}(L))=d_L^2(z_{\rm ref})\times L/(L_{\rm ref}/k),
\end{equation}
where $k$ is the ratio of the reference source flux to the detector
sensitivity (see Table \ref{table:sen}).

\begin{table}
\centering
\caption{Number of TDEs per year above MM detection thresholds}
\begin{tabular}{cccccc}
\hline \hline
    & LMT/AzTEC & ALMA \\
\hline
On-axis  & 0.6 & 10 \\
Off-axis ($a_{\rm off}^4=10^{-4}$) & 13 & 220 \\
Off-axis ($a_{\rm off}^4=10^{-3}$) & 130 & 2230 \\
Off-axis ($a_{\rm off}^4=10^{-5}$) & 1.3 & 22 \\
\hline
\hline
\end{tabular}
\label{table:num}
\end{table}

Fig. \ref{fig:evrate} shows the event rate density as a function of the
peak bolometric luminosity and redshift, for on-axis (left) and off-axis
(right) jetted TDEs, respectively. The curves overplotted show the
MM detection threshold (or $z_{\rm max}$ for a given $L$) above which
the sources are detectable, assuming the same peak bolometric luminosity
to peak MM flux ratio of Sw J1644+57 (IGR J12580+0134) for on-axis
(off-axis) sources. Integrating the regions above the curves shown in
Fig. \ref{fig:evrate} gives the event rate, as listed in Table
\ref{table:num}. In reality, one also needs to account for the trigger
efficiency of TDEs \citep{2014MNRAS.437..327K,2015ApJ...803...36D},
as well as the visibility by the MM telescopes\footnote{For example,
an event can only be detected about half years later when its flux
declines much from the peak flux, if it happens to occur in the daytime
for the LMT. This may not be a big problem for the ALMA.}.

\begin{figure}
\centering
\includegraphics[width=\columnwidth]{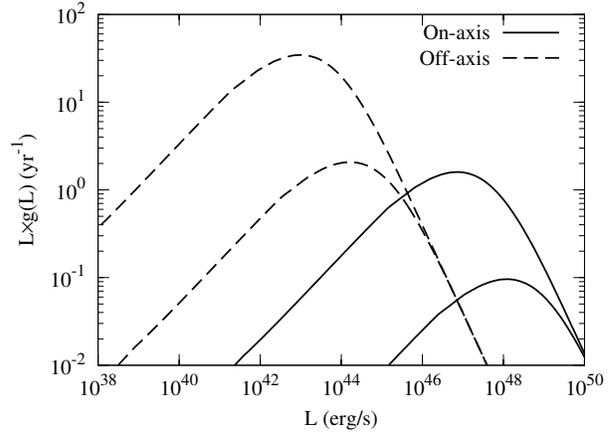}
\caption{Luminosity distribution $L\times g(L)$ of detectable TDEs as
defined in Eq. (\ref{eq:gl}). Solid and dashed lines are for on-axis
and off-axis ($a_{\rm off}^4=10^{-4}$) TDEs, respectively. For each type,
the higher one is for the ALMA, and the lower one is for the LMT.
}
\label{fig:gL}
\end{figure}

We define the redshift integral part of Eq. (\ref{eq:num}) as
\begin{equation}
g(L)=\frac{\Omega}{4\pi}\int_0^{z_{\rm max}(L)}dz\frac{\rho(L,z)}{1+z}
\frac{dV}{dz},
\label{eq:gl}
\end{equation}
which gives the luminosity distribution of detectable TDEs. The distributions
$L\times g(L)$ for the LMT and ALMA are shown in Fig. \ref{fig:gL}, for on-
and off-axis TDEs. The high-$L$ end of the distributions follows the intrinsic
luminosity function $L^{-2}$ as shown in Eq. (\ref{eq:on}), which suggests
an almost full coverage of such events. The low luminosity ones can only be
detected within a limited volume, and thus their numbers decrease again.
The results show that for on-axis TDEs those have luminosities $L\sim10^{47}
-10^{48}$ erg/s are most probably be detected, while for off-axis TDEs the
most probable luminosites are $L\sim10^{43}-10^{44}$ erg/s. Note that
these values depend on the assumption of $a_{\rm off}$.

\begin{figure}
\centering
\includegraphics[width=\columnwidth]{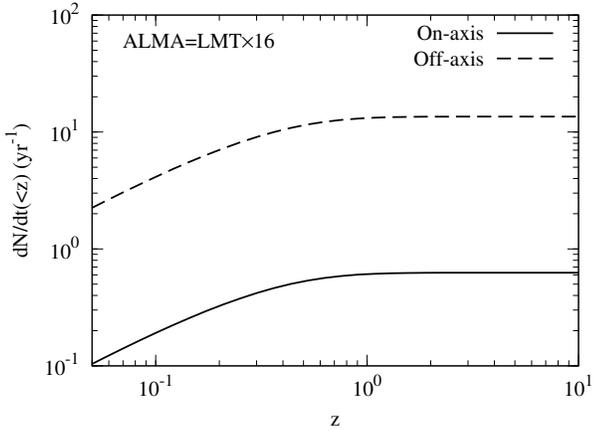}
\caption{Cumulative rate of detectable TDEs $dN/dt(<z)$ versus redshift,
for the LMT. The results for the ALMA can be obtained through a scaling
upwards by a factor of 16.
}
\label{fig:Nz}
\end{figure}

Finally we show the redshift distribution of the cumulative rate of
detectable TDEs in Fig. \ref{fig:Nz}. We find that most of the detectable
TDEs will be within $z\sim1$. This is expected because in the distant
Universe only high luminosity sources are detectable, which are rare due
to their steep luminosity function. The redshift distribution of SMBHs,
$f(z)$, suppresses the event rate of TDEs at high redshifts further.

\section{Potential scientific yields}

We briefly discuss the possible physical insights into the TDE jets and
related problems that could be addressed via the MM and multi-wavelength
observations.

\subsection{Jetted TDE statistics}
At present the jetted TDE sample is too small to allow for sensible
statistical studies of them. The radio follow-ups of TDEs suggest
a jetted TDE fraction of the order of $\sim10\%$ \citep{2013ApJ...763...84B,
2013A&A...552A...5V}. The fraction of the currently detected jetted to
the whole TDE sample implies a similar value. However, this fraction can
only be estimated reliably with a significantly enlarged sample.
More jetted TDE events will also enable us to study the diversity of
the jets, as well as their correlations with related physical parameters
such as the SMBH spin \citep{2011ApJ...740L..27L} and the properties of
the disrupted objects. The MM observations, which
should be efficient to catch the jet emission early, will effectively
increase the number of detections of jetted TDEs in the near future.
A well-defined sample will help us to understand the nature of TDEs
and their associated jet formation.

\subsection{Jet dynamics and CNM density structure}
The MM monitoring of the jet emission, especially at the early stages which
are relatively difficult to be probed in the radio bands due to the optical
thickness, is important to understand the jet dynamical evolution.
The early emission gives effective constraints on the jet parameters, such
as the initial Lorentz factor, the deceleration time, and the total kinetic
energy, while the late-time emission is more sensitive to the density
structure of the CNM. The combination of the multi-epoch and multi-wavelength
observations will enable us to develop a comprehensive model for the evolution
of the jets when they propagate in the CNM, tightly constraining parameters
related to the particle acceleration ($\nu_m$, $p$) and cooling ($\nu_c$),
as well as the optical depth of the emission region ($\nu_a$) (see, e.g.,
\S~3.2 for Sw J1644+57 as an example).

The CNM density structure is a key manifestation of the SMBH accretion
process, which still remains rather uncertain. The monitoring of jetted
TDEs can provide a powerful (if not unique) prober of its structure.
In particular, the late-time light curves (in Sedov-Taylor phase) can
reveal solely the propagation effect of the jets in the CNM. The decline
slope of the emission depends sensitively on the density profile of the
CNM, as well as the spectral index of accelerated electrons. The latter
can be determined through measurements of the in-band spectrum and/or
multi-wavelength SED. Thus the CNM density profile can be readily obtained
\citep{2012ApJ...748...36B,2016ApJ...819L..25A}.

\subsection{Magnetic field in the CNM}
Another fundamental ingradient related to the SMBH accretion and feedback,
jet production, and cosmic ray transportation is the magnetic field in
the CNM. TDE jets provide again a unique tool to probe the magnetic field
structure which is generally very difficult to be studied except for the
nuclei of the Milky Way and possibly a few nearby galaxies \citep[e.g.,][]
{2007ApJ...654L..57M,2013Natur.501..391E,2015Sci...350.1242J,
2013IAUS..294..213H}. The rotation measure of the radio-MM emission from
the TDE jets when they propagate in the CNM can be used to probe the
magnetic field structure in the host nuclei \citep{2007ApJ...654L..57M,
2013Natur.501..391E,2015ApJ...798...22L}. The Faraday rotation angle
is strongly wavelength dependent ($\propto\lambda^2$).
For a typical rotation measurement of $\sim10^5$ rad m$^{-2}$
\citep{2007ApJ...654L..57M,2013Natur.501..391E}, MM polarization
observations can be used to give an unambiguous measure of the rotation
angle of the orders of tens of degrees. Such measurements are important
for determining the intrinsic polarization angles of the jets, as well
as the magnetic field structure of the CNM.

\subsection{Acceleration of ultra-high energy cosmic rays}
TDEs are candidate sources of cosmic rays, probably up to ultra-high
energies \citep[UHECRs;][]{2009ApJ...693..329F,2014arXiv1411.0704F}.
The maximum achievable energy of a proton in an acceleration site is
given by the product of the size and magnetic field of the source:
$E_{\rm max}\sim qBR\Gamma$ \citep{1984ARA&A..22..425H,
2009ApJ...693..329F}. For TDE jets, $R$ is on the order of pc, $B$
can be up to Gauss, and hence $E_{\rm max}\sim 10^{20}$ eV. Assuming
an event rate density of $\sim3\times10^{-11}$ Mpc$^{-3}$ yr$^{-1}$
and an isotropic, bolometric energy released per event of
$\sim10^{54}$ erg for on-axis jetted TDEs\footnote{Off-axis jetted
TDEs do not contribute additionally to this estimate because the
beaming factors in estimating the event rate density and the
realistic energy cancels out with each other.}, the energy injection
rate in radiation would be about $3\times10^{43}$ erg Mpc$^{-3}$
yr$^{-1}$ \citep{2016PhRvD..93h3005W}. The energy injection rate in
cosmic rays could be even higher than in the electro-magnetic
radiation, and could be comparable to that required to explain the
observed UHECR flux \citep{1995ApJ...452L...1W}. Therefore TDEs
could indeed be promising candidate sources of UHECRs.

\section{Summary}

%TDEs are unique objects to probe the inactive, majorate population of
%SMBHs and the central environments of their host galaxies, as well as the
%associated jet launching processes. They are also attractive candidates to
%accelerate UHECRs. One of the greatest advantages of TDEs is their long
%time scales (up to years) which make the follow-up observations be easy
%to carry out, even at the earliest stages of such events. Timely and
%adequate monitorings of TDEs in multi-bands are very important to
%understand the nature of TDEs themselves and the interplay between
%SMBHs and the CNM (e.g., the density and magnetic field profiles).

Jetted TDEs can be a uniquely powerful tool to probe the overall population
of SMBHs and the CNM of their host galaxies, as well as the physics of
jet formation and particle acceleration. In this work we have investigated
the potential of observing jetted TDEs in the MM bands. Such observations
are optimal to capture the launching and earliest evolution of the jets,
which are crucial to understanding the jet physics and the jet-CNM
interaction, due to the transparency of the MM emission. Our findings
are as follows.

\begin{itemize}

\item
With the Planck survey data, we detect an MM counterpart of IGR
J12580+0134, a nearby TDE discovered in December 2010. The MM
counterpart is positionally coincident with the TDE and most
importantly shows a consistent variability behaviour. The flux
densities of the HFI measurements, together with the late time data,
are consistent with the expectation of a time-dependent jet
evolution model. This detection illustrates the feasibility of
detecting early MM emission from jetted TDEs, even with an off-axis
geometry.

\item We have conducted a systematic search of potential MM
counterparts of known TDEs, based on the flux density variability
detected in the PCCS1 and PCCS2 catalogues. No detection is found
for the other $\sim20$ TDE candidates which occured during the
Planck's operation. A cross-correlation between variable Planck
sources with nearby galaxies may allow for detecting new TDE
candidates.

\item
We model the multi-wavelength and multi-epoch emissions from jetted TDEs,
either on-axis (like Sw J1644+57) or off-axis (like IGR J12580+0134),
with realistic jet-CNM interaction models. We find that MM observations
of TDEs, especially at early evolutionary stages (e.g., within one month
after triggering), are sensitive to probe the spectral index and minimum
energy of accelerated electrons.

\item
Taking Sw J1644+57 and IGR J12580+0134 as examples of on- and off-axis
jetted TDEs, we investigate the potential of observing them with operating
MM facilities such as the LMT and ALMA. We find that both types of events
could be detectable up to redshifts of $\sim1$ (2) by the LMT (ALMA), for
a disrupted star mass of $\sim$M$_\odot$. With a systematic following-up
monitoring program, the LMT (ALMA) could detect as many as $\sim0.6$ (13)
on- and 10 (220) off-axis TDEs per year, which are, however, limited by
the trigger rate in optical/X-ray surveys.

\item
Extensive new MM observations, together with other multi-wavelength
follow-ups, can lead to major advances in our understanding of the jet
dynamics, the density and magnetic field structures of the CNM, and the
origin of UHECRs --- all are important issues in high-energy astrophysics.

\end{itemize}

We conclude that the development of a comprehensive MM follow-up observing
program of TDEs is both highly desirable and timely. Such a program will
enable us to take advantage of the rapidly improved TDE triggering
capabilities provided by existing time-domain astronomical facilities such
as Swift \citep{2004ApJ...611.1005G}, Fermi/GBM \citep[Fermi Gamma-ray Burst
Monitor;][]{2009ApJ...702..791M}, ASASSN (All-Sky Automated Survey for
SuperNovae\footnote{http://www.astronomy.ohio-state.edu/~assassin/index.shtml}),
Pan-STARRS \citep[Panoramic Survey Telescope \& Rapid Response System;]
[]{2010SPIE.7733E..0EK}, and DES \citep[Dark Energy Survey;][]
{2005IJMPA..20.3121F}, as well as upcoming eROSITA \citep[extended ROentgen
Survey with an Imaging Telescope Array;][]{2012arXiv1209.3114M},
EP \citep[Einstein Probe;][]{2015arXiv150607735Y},
ZTF \citep[Zwicky Transient Facility;][]{2014htu..conf...27B},
and LSST \citep[Large Synoptic Survey Telescope;][]
{2009arXiv0912.0201L}. MM follow-ups will then play a significant role in
advancing the understanding of TDEs, the population of low luminosity SMBHs,
and many related astrophysical phenomena and processes discussed above.

\section*{Acknowledgments}

We thank Timonthy J. Pearson, Jorg P. Rachen, and Yu-Ping Tang for
valuable suggestions and comments on the Planck data analysis, and
Xiang-Yu Wang for helpful discussion on the modeling of TDEs.
%This work is supported by 973 Program under Grant No. 2013CB837000, and
%by National Natural Science Foundation of China under Grant No. 11105155.

\bibliographystyle{mn2e}
\bibliography{refs}

\end{document}